\documentclass[sigconf]{acmart}

\usepackage{bigstrut}
\usepackage{booktabs}
\usepackage{multirow}
\usepackage{microtype}
\usepackage{graphicx}
\usepackage{subfigure}
\usepackage{subcaption}
\usepackage{overpic}
\usepackage{balance}
\usepackage{amsthm}
\usepackage{array}
\usepackage{clrscode}
\usepackage{multicol}
\usepackage{float}
\usepackage{newfloat}
\usepackage{color}
\usepackage{xcolor}
\usepackage{amsopn}
\usepackage{mathrsfs}
\usepackage{mathtools}
\usepackage{amsmath}
\usepackage{arydshln}
\usepackage{blkarray}
\usepackage{enumerate}
\usepackage{courier}
\usepackage{rotating}
\usepackage{bm}
\usepackage{wrapfig}
\usepackage{ragged2e}
\usepackage[misc]{ifsym}
\usepackage{makecell}
\usepackage{adjustbox} %
\usepackage{enumitem}
\usepackage{hyperref}
\usepackage{url}
\usepackage{xspace}
\usepackage{comment}
\usepackage{changepage}
\usepackage{algorithm}
\usepackage{algpseudocode} 
\usepackage{algorithmicx} 
\usepackage[ruled, vlined, nofillcomment, linesnumbered, algo2e]{algorithm2e}
\usepackage{tcolorbox}
\usepackage{balance}

\newcommand{\ie}{\emph{i.e.,}\xspace}
\newcommand{\eg}{\emph{e.g.,}\xspace}

\newcommand{\ignore}[1]{}

\AtBeginDocument{%
  }
\copyrightyear{2026}
\acmYear{2026}
\setcopyright{cc}
\setcctype{by-nc-nd}
\acmConference[KDD 2026] {Proceedings of the 32nd ACM SIGKDD Conference on Knowledge Discovery and Data Mining V.2}{August 9--13, 2026}{Jeju Island, Republic of Korea.}
\acmBooktitle{Proceedings of the 32nd ACM SIGKDD Conference on Knowledge Discovery and Data Mining V.2 (KDD 2026), August 9--13, 2026, Jeju Island, Republic of Korea}
\acmISBN{979-8-4007-2259-2/2026/08}
\acmDOI{10.1145/3770855.3818451}

\begin{document}

\title{MSN: A Memory-based Sparse Activation Scaling Framework for Large-scale Industrial Recommendation}


\author{Shikang Wu}
\email{wushikang@bytedance.com}
\authornote{These authors contributed equally.}
\affiliation{
    \institution{ByteDance Search}
    \city{Beijing}
    \country{China}
}

\author{Hui Lu}
\email{luhui.xx@bytedance.com}
\authornotemark[1]
\affiliation{
    \institution{ByteDance AML}
    \city{Hangzhou}
    \country{China}
}

\author{Jinqiu Jin}
\email{jinjinqiu.02@bytedance.com}
\authornotemark[1]
\affiliation{
    \institution{ByteDance Search}
    \city{Beijing}
    \country{China}
}

\author{Zheng Chai}
\email{chaizheng.cz@bytedance.com}
\authornotemark[1]
\affiliation{
    \institution{ByteDance AML}
    \city{Hangzhou}
    \country{China}
}

\author{Shiyong Hong}
\email{hongshiyong.66@bytedance.com}
\affiliation{
    \institution{ByteDance Search}
    \city{Beijing}
    \country{China}
}

\author{Junjie Zhang}
\email{zhangjunjie.leo@bytedance.com}
\affiliation{
    \institution{ByteDance Search}
    \city{Beijing}
    \country{China}
}

\author{Shanlei Mu}
\email{mushanlei@bytedance.com}
\affiliation{
    \institution{ByteDance Search}
    \city{Beijing}
    \country{China}
}

\author{Kaiyuan Ma}
\email{makaiyuan.m@bytedance.com}
\affiliation{
    \institution{ByteDance AML}
    \city{Beijing}
    \country{China}
}

\author{Tianyi Liu}
\email{liutianyi.2024@bytedance.com}
\affiliation{
    \institution{ByteDance AML}
    \city{Beijing}
    \country{China}
}

\author{Yuchao Zheng}
\email{zhengyuchao.yc@bytedance.com}
\affiliation{
    \institution{ByteDance AML}
    \city{Hangzhou}
    \country{China}
}

\author{Zhe Wang}
\authornote{Corresponding Author.}
\email{zhewang.tim@gmail.com}
\affiliation{
    \institution{ByteDance Search}
    \city{Beijing}
    \country{China}
}

\author{Jingjian Lin}
\email{linjingjian000@gmail.com}
\affiliation{
    \institution{ByteDance Search}
    \city{Beijing}
    \country{China}
}

\renewcommand{\shortauthors}{Shikang Wu et al.}

\begin{abstract}

Scaling deep learning recommendation models is an effective way to improve model expressiveness. Existing approaches often incur substantial computational overhead, making them difficult to deploy in large-scale industrial systems under strict latency constraints. Recent sparse activation scaling methods, such as Sparse Mixture-of-Experts, reduce computation by activating only a subset of parameters, but still suffer from high memory access costs and limited personalization capacity due to the large size and small number of experts. To address these challenges, we propose \textbf{MSN}, a memory-based sparse activation scaling framework for recommendation models. MSN dynamically retrieves personalized representations from a large parameterized memory and integrates them into downstream feature interaction modules via a memory gating mechanism, enabling fine-grained personalization with low computational overhead. To enable further expansion of the memory capacity while keeping both computational and memory access costs under control, MSN adopts a Product-Key Memory (PKM) mechanism, which factorizes the memory retrieval complexity from linear time to sub-linear complexity. In addition, normalization and over-parameterization techniques are introduced to maintain balanced memory utilization and prevent memory retrieval collapse. We further design customized Sparse-Gather operator and adopt the AirTopK operator to improve training and inference efficiency in industrial settings. Extensive experiments demonstrate that MSN consistently improves recommendation performance while maintaining high efficiency. Moreover, MSN has been successfully deployed in the Douyin Search Ranking System, achieving significant gains over deployed state-of-the-art models in both offline evaluation metrics and large-scale online A/B test.
\end{abstract}

\ccsdesc[500]{Information systems~Recommender systems}
\keywords{Large Recommendation Models, Memory Network, Sparse Activation Scaling}

\maketitle

\section{Introduction}
\label{sec:intro}

Recommender systems are essential for online platforms in helping users discover relevant content efficiently. With recent developments in deep learning, recommendation models have shifted from shallow models to Deep Learning Recommendation Models (DLRMs), which aims to enhance model capacity by designing effective feature interaction architectures~\cite{covington2016deep,guo2017deepfm,wang2021dcn,zhang2024wukong,zhou2018deep}. Notably, recent advances in large language models (LLMs) demonstrate that increasing model parameters can effectively improve performance, inspiring growing interest in scaling DLRMs to boost model capacity. Most existing recommendation model scaling strategies either expands sparse embedding tables~\cite{guo2024embedding,pan2024ads} or enlarges feature interaction modules~\cite{zhu2025rankmixer,lai2025exploring}. Despite effectiveness in improving recommendation quality, they inevitably introduce substantial computational overhead. As industrial recommender systems have strict low latency and high throughput demands, traditional scaling strategies often face limitation in practical deployment. Thus, \emph{how to effectively scale recommendation models without compromising efficiency remains a key challenge}.

To solve this problem, recent works adopt sparse activation scaling techniques to dynamically activate only a subset of model parameters. This allows growing model capacity while incurring marginal computational overhead. For example, Sparse Mixture-of-Experts (SMoE) architectures~\cite{riquelme2021scaling} only selectively activate several experts for different samples, so that the computation cost for each instance remains roughly constant with increasing model's total parameters. However, such SMOE-based scaling approaches still have the following limitations. Firstly, existing SMoE-based methods primarily focus on reducing computational cost, while largely overlooking the memory access cost during both training and inference. Specifically, in a standard SMoE layer with hidden size $d$, each activated expert corresponds to a full Feed-Forward Network (FFN). This results in a memory access cost of $\mathcal{O}(d^2)$, which becomes a dominant performance bottleneck under bandwidth-limited inference settings. Secondly, while SMoE architectures enable fine-grained user modeling through customized expert activation~\cite{zhu2025rankmixer,qi2025mtmixatt}, the number of activated experts is inherently constrained. As established earlier, each expert's $\mathcal{O}(d^2)$ parameter size would restrict the total number of experts under practical hidden size $d$ scaling. Consequently, each user can only activate a small subset from the limited expert pool, which results in limited personalization capability and suboptimal recommendation performance. Thus, there calls for a sparse activation scaling method that is more memory-efficient and capable of personalization.

Rather than sparse computation like SMoE, some recent works in LLMs achieve linear memory complexity via sparse parameter retrieval. Inspired by this, we can design a memory-centric architecture~\cite{he2024mixture,zhong2024memorybank,huangultra} for efficient sparse scaling in recommendation models. Specifically, instead of routing each interaction through a limited set of shared experts, we can dynamically retrieve a value representation from a large parameterized memory for each user. This design enables the model capacity to scale with the size of the memory table and allows user-specific information to be encoded through lightweight memory access. While this approach is appealing, it presents several major challenges when employed at the industrial recommendation systems.  Firstly, as recommendation models typically rely on complex feature interaction modules to capture fine-grained user preferences, it remains a non-trivial challenge to incorporate the retrieved memory into downstream modeling. Secondly, with billions of users in large-scale recommender platforms, maintaining explicit memory representations for individual users is infeasible due to prohibitive parameter and storage costs. Thirdly, memory retrieval can induce highly non-uniform access patterns, leading to load imbalance and degraded efficiency during training and inference.

To address the above challenges, we propose \textbf{MSN} (\textbf{M}emory \textbf{S}caling \textbf{N}etwork), a novel memory-based sparse activation scaling framework for recommendation models. The core idea is employing an efficient retrieval mechanism to dynamically fetch personalized memory representations from a large-scale parameterized storage. Building upon this, we introduce a \emph{Memory Gating Mechanism} to effectively apply the retrieved memory values on downstream modules, where the retrieved memory acts as a modulation signal to the original feature representations for better personalization. To achieve scalable while efficient memory retrieval in the key-value memory network, we adopt the Product-Key Memory (PKM)~\cite{lample2019large} to factorize the original memory's index space into two key-subspaces for reducing computational complexity. Furthermore, to ensure \emph{Stable and Balanced Memory Optimization}, we apply Layer Normalization~\cite{ba2016layer} and Learning Rate Warm-up on both query and key vectors, while employing an Over-Parameterization strategy on original key embeddings. At last, we introduce the customized \emph{Sparse-Gather} operator and adopt AirTopK~\cite{zhang2023parallel} operator for \emph{efficient training and deployment} in large-scale industrial systems. 

In summary, MSN enables model capacity to scale sub-linearly with the memory table size, while introducing only lightweight retrieval and fusion operations at the computational level. This substantially improves model expressiveness while maintaining inference efficiency. It is noteworthy that MSN can also serve as a plug-in model-agnostic scaling module on various backbones, which is shown to consistently achieve superior performance. Specifically, MSN has been deployed on Douyin Search Ranking System based-on existing SMoE-based models, which has achieved significant offline and online performance improvement.


\section{Related Work}\label{sec:related}

In this section, we briefly review recent works about \emph{Deep Learning Recommendation Models}, \emph{Scaling up Recommendation Models}, and \emph{Memory Modules for LLM Scaling}.

\vspace{3pt}
\noindent${\bullet}\ $\textbf{Deep Learning Recommendation Models. }
In recent years, Deep Learning Recommendation Models (DLRMs) have become core components of large-scale industrial systems, aiming to handle heterogeneous features by designing effective feature interaction modules~\cite{zhang2019deep}. Representative models such as Wide\&Deep~\cite{widedeep}, DeepFM~\cite{guo2017deepfm}, NFM~\cite{nfm}, xDeepFM~\cite{lian2018xdeepfm}, and HOFM~\cite{hofm} typically combine a linear component that captures low-order, memorization-friendly patterns with a neural network that models high-order interactions, thereby unifying low- and high-order feature learning within a single framework. Building on this idea, models including DCN~\cite{dcn}, DCNv2~\cite{wang2021dcn}, DeepCross~\cite{deepcross}, FiGNN~\cite{fignn}, and FinalMLP~\cite{mao2023finalmlp} further increase modeling capacity by introducing specialized architectures, such as cross networks or graph-based structures, to explicitly learn complex high-order combinatorial relationships beyond standard feedforward layers. More recently, attention mechanisms have been widely used to model dynamic and asymmetric interactions, where methods such as DIN~\cite{zhou2018deep}, DIEN~\cite{zhou2019deep}, AutoInt~\cite{song2019autoint}, Hiformer~\cite{gui2023hiformer}, TIN~\cite{tin}, and DPN~\cite{dpn} apply attention to selectively weight features or user behaviors, enabling context-aware and non-uniform feature interaction modeling. These DLRM architectures have served as the foundation for scaling up recommender models in recent years.

\vspace{3pt}
\noindent${\bullet}\ $\textbf{Scaling up Recommendation Models. }
Motivated by the success of LLMs where increasing model parameters often leads to better performance, there has been growing interest in scaling DLRMs to better exploit large-scale data~\cite{zhang2024wukong,zhai2024actions,shen2024optimizing,han2025mtgr,wang2025scaling,li2025realizing,lv2025marm,lai2025exploring,qi2025mtmixatt}. Prior studies have explored different approaches to increase model capacity while improving recommendation accuracy. For example, Wukong~\cite{zhang2024wukong} introduces a stacked factorization machine architecture and a synergistic upscaling strategy, which enables the model to capture diverse, any-order feature interactions through increasing its depth and width. HSTU~\cite{zhai2024actions} scales up DLRMs by reformulating recommendation as a generative sequential transduction problem and introducing a Transformer-style architecture optimized for high-cardinality streaming data. MTGR~\cite{han2025mtgr} builds on generative recommender architectures to preserve rich DLRM cross features while introducing user-level compression and system-level optimizations, enabling efficient training and inference at tens-to-hundreds times higher computational complexity. RankMixer~\cite{zhu2025rankmixer} designs a hardware-aware unified feature interaction architecture that replaces quadratic self-attention with an efficient multi-head token mixing module, and scales to billion parameters via a SMoE variant, achieving a 100x parameter increase while maintaining inference latency. In this work, we explore a plug-and-play memory network-based framework for scaling up DLRMs.

\vspace{3pt}
\noindent${\bullet}\ $\textbf{Memory Modules for LLM Scaling. }
Recently, there have been surging works which dedicates on introducing memory modules into LLM backbones to enhance their modeling capacities and scaling up performances~\cite{zhao2023survey}. For instance, LongMem~\cite{wang2023augmenting} augments a frozen backbone with a decoupled long-term memory and adaptive retrieval mechanism, enabling efficient utilization of unlimited-length historical context without increasing computational complexity. UltraMem~\cite{huangultra} and UltraMemV2~\cite{huang2025ultramemv2} replace sparse expert activation with efficient memory-layer architectures that achieve MoE-level performance under equivalent compute while substantially reducing memory access overhead. Memory+~\cite{bergesmemory} and STEM~\cite{sadhukhan2026stem} replaces the FFN up-projection with token-indexed embedding modules, which decouples parameter growth from per-token FLOPs and enables test-time capacity scaling as sequence length increases. Engram~\cite{cheng2026conditional} scales LLMs by introducing conditional memory as a complementary sparsity axis to MoE, enabling O(1) knowledge lookup that shifts static information from computation to memory and yields better performance. In this work, we aim to employ the sparse memory network into DLRMs for better model scaling.

\section{Methodology}\label{sec:method}

In this section, we elaborate the proposed MSN framework for scaling up recommendation models.

\subsection{Problem Definition}\label{subsec:problem}

We focus on the click-through rate (CTR) prediction task, which aims at using abundant features to estimate the probability of user clicks over items. We denote the training samples as $\mathcal{D}=\{(\bm{x},y)\}$, where $\bm{x}$ is a sparse feature vector (such as user profiles, item descriptions and contextual signals), and $y\in\{0,1\}$ is the target label (\ie click or not) collected from historical logs. 

The objective is to estimate the click probability $P(y=1|\bm{x})$ using a recommendation model $f_{\bm{\theta}}(\bm{x})$ with a specific architectural design, where $\bm{\theta}$ denotes the model parameters. Recent progress in deep learning–based recommendation models (DLRMs) shows that, with properly designed model architectures, scaling model capacity by increasing the number of parameters is an effective way to improve CTR prediction performance.
The parameter $\bm{\theta}$ is then optimized over the historical clicks $\mathcal{D}$ by minimizing the standard binary cross-entropy loss
\begin{equation*}
    \mathcal{L}=-\frac{1}{|\mathcal{D}|}\sum\limits_{(\bm{x},y)\in\mathcal{D}}\left(y\log \sigma\left(\hat{y}\right)+\left(1-y\right)\log\left(1-\sigma\left(\hat{y}\right)\right)\right),
\end{equation*}
where $|\mathcal{D}|$ denotes the number of training samples and $\sigma(\cdot)$ denotes the \textit{sigmoid} function to normalize the prediction.

\subsection{Memory Scaling Network Overview}
\label{subsec:overview}

\begin{figure*}[t]
\centering
\includegraphics[width=0.65\textwidth]{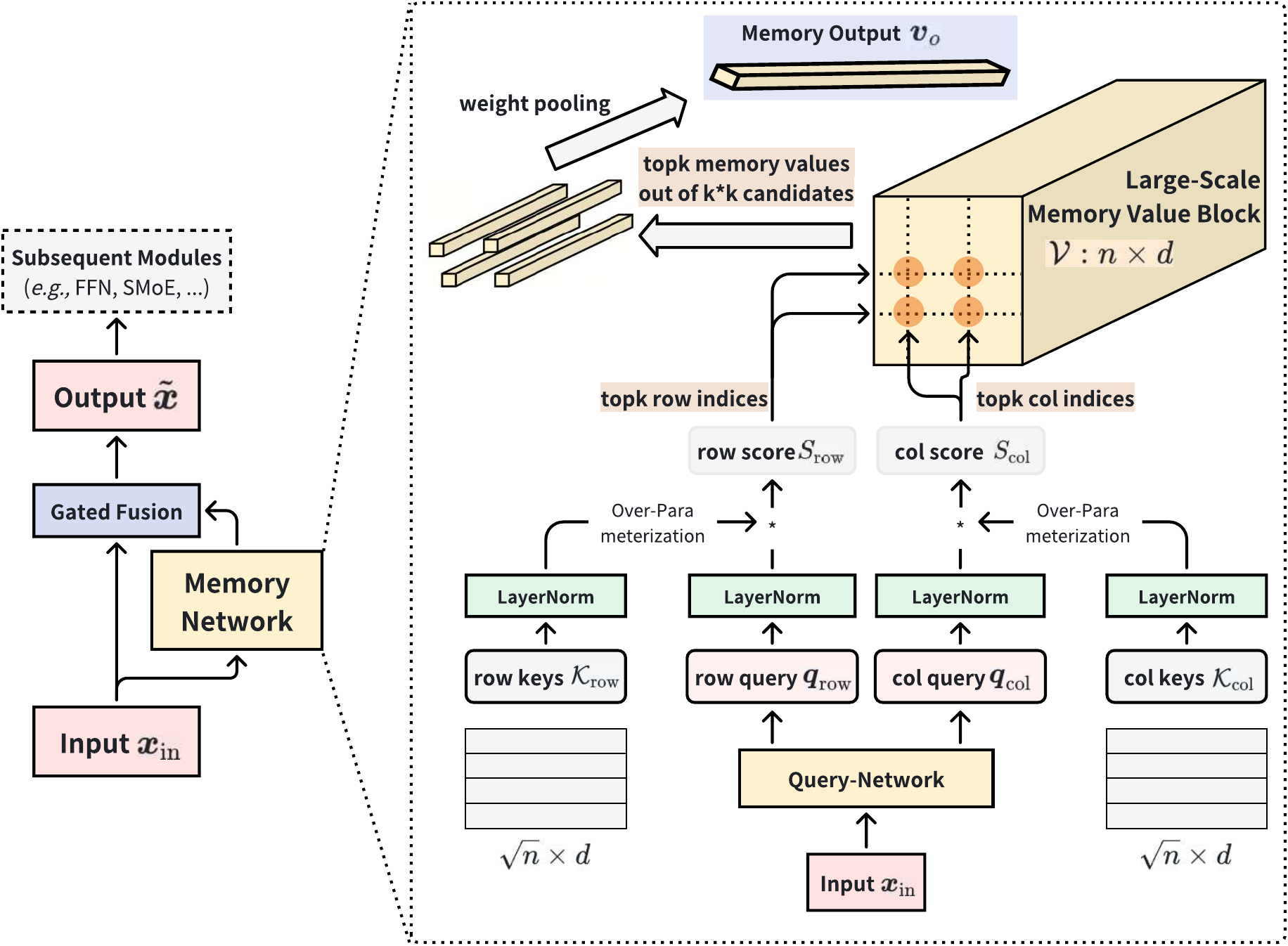}
\vspace{-5pt}
\caption{Overall framework of MSN.}
\label{fig:overall}
\vspace{-5pt}
\end{figure*}

In this work, we aim to develop a sparse activation scaling framework for recommendation models that achieves strong personalization while remaining memory-efficient. Unlike prior sparse activation scaling approaches such as SMoE, which primarily rely on sparse computation, we explore an alternative paradigm based on \emph{sparse parameter retrieval} by introducing a large-scale Memory Network for recommendation modeling. The overall framework of our proposed Memory Scaling Network (MSN) is shown as Figure~\ref{fig:overall}. 

Specifically, instead of routing each input through a small set of shared experts, MSN dynamically retrieves a value representation from a parameterized memory conditioned on the input. This design allows the model capacity to scale with the size of the memory table, while only incurring lightweight retrieval overhead. As a result, the expressive power of the model can be significantly increased without violating the strict latency constraints of industrial recommender systems. In addition, similar to observations in recent memory-based LLM architectures, this design can also enable seamless integration with existing modules (such as SMoE blocks) and consistently achieves superior performance.

\subsection{Memory Retrieval with Product Keys}
\label{subsec:pkm}

\subsubsection{Motivation}
As illustrated before, the core idea of MSN is to build
a large-scale memory value block $\mathcal{V}\in \mathbb{R}^{n \times d}$, and retrieve Top-$k$ relevant value representations conditioned on the input, enabling fine-grained personalization while improving generalization. In a standard key–value memory network, memory retrieval requires computing similarities between the query input and all memory keys. However, the retrieval complexity grows linearly with memory size $n$, making it difficult to scale to large memory capacities.

To address this issue, we adopt the \emph{Product Key Memory (PKM)} mechanism~\cite{lample2019large}. Specifically, we decompose the global memory index space into two sub-key spaces with learnable parameters $\mathcal{K}_{\text{row}}, \mathcal{K}_{\text{col}} \in \mathbb{R}^{\sqrt{n} \times d}$. During retrieval, the query vector independently performs Top-$k$ search over both sub-key matrices to obtain preliminary row and column indices. Afterwards, each memory value is implicitly indexed by the Cartesian product of the two selected sub-key sets. This factorization transforms full-space retrieval into two independent subspace retrievals, largely reducing the computational complexity from $\mathcal{O}(nd)$ to $\mathcal{O}(\sqrt{n}d)$.

\subsubsection{Memory Indices and Values Retrieval}
Given the query input $\bm{x}_{\text{in}}$, a query network firstly generates two subspace-specific queries $\bm{q}_{\text{row}},\bm{q}_{\text{col}}\in\mathbb{R}^{d}$ as follows
\begin{equation*}
\bm{q}_{\text{row}},\bm{q}_{\text{col}} = \text{Query-Network}(\bm{x}_{\text{in}}).
\end{equation*}
Then the similarity scores between queries and sub-keys are computed via dot products
\begin{equation}\label{eq:S_old}
S_{\text{row}} = \mathcal{K}_{\text{row}}\bm{q}_{\text{row}},\  S_{\text{col}} = \mathcal{K}_{\text{col}}\bm{q}_{\text{col}}.
\end{equation}
Based on subspace similarity scores $S_{\text{row}},S_{\text{col}}\in\mathbb{R}^{\sqrt{n}}$, we firstly select their corresponding Top-$k$ indices independently, and their combinations with candidate size $k^2$ yield the final Top-$k$ indices. We visualize the whole process in Figure \ref{fig:overall}, which can be mathmatically expressed as follows
\begin{equation*}
\begin{aligned}
\mathcal{I}&=\mathop{\text{argtop}k}\limits_{i\in[0,\sqrt{n}-1]}\,S_{\text{row}},\quad \mathcal{J}=\mathop{\text{argtop}k}\limits_{j\in[0,\sqrt{n}-1]}\,S_{\text{col}}, \\
\mathcal{K}&=\mathop{\text{argtop}k}\limits_{(i,j)\in\mathcal{I}\times\mathcal{J}}\ S_{\text{row}}[i]+S_{\text{col}}[j].
\end{aligned}
\end{equation*}

After obtaining the Top-$k$ sub-key index pairs $(i,j)$, we retrieve the corresponding memory values based on the computed indices. These retrieved memory values are then aggregated into the memory layer output $\bm{v}_o\in\mathbb{R}^d$ via a softmax-weighted sum 
\begin{equation}\label{eq:vo_mem}
\begin{aligned}
\bm{v}_o&=\sum_{(i,j)\in\mathcal{K}}S[i,j]\cdot\mathcal{V}[(\sqrt{n}-1)\cdot i+j], \\
\text{where }S[i,j]&=\frac{S_{\text{row}}[i]+S_{\text{col}}[j]}{\sum_{(i,j)\in\mathcal{K}}(S_{\text{row}}[i]+S_{\text{col}}[j])}.
\end{aligned}
\end{equation}

The above product key factorization design enables the memory size to scale to hundreds of thousands of entries while maintaining feasible computational overhead.

\subsubsection{Structural Comparisons with stardard FFN and SMoE}

\begin{figure}[t]
\centering
\vspace{-5pt}
\includegraphics[width=0.83\columnwidth]{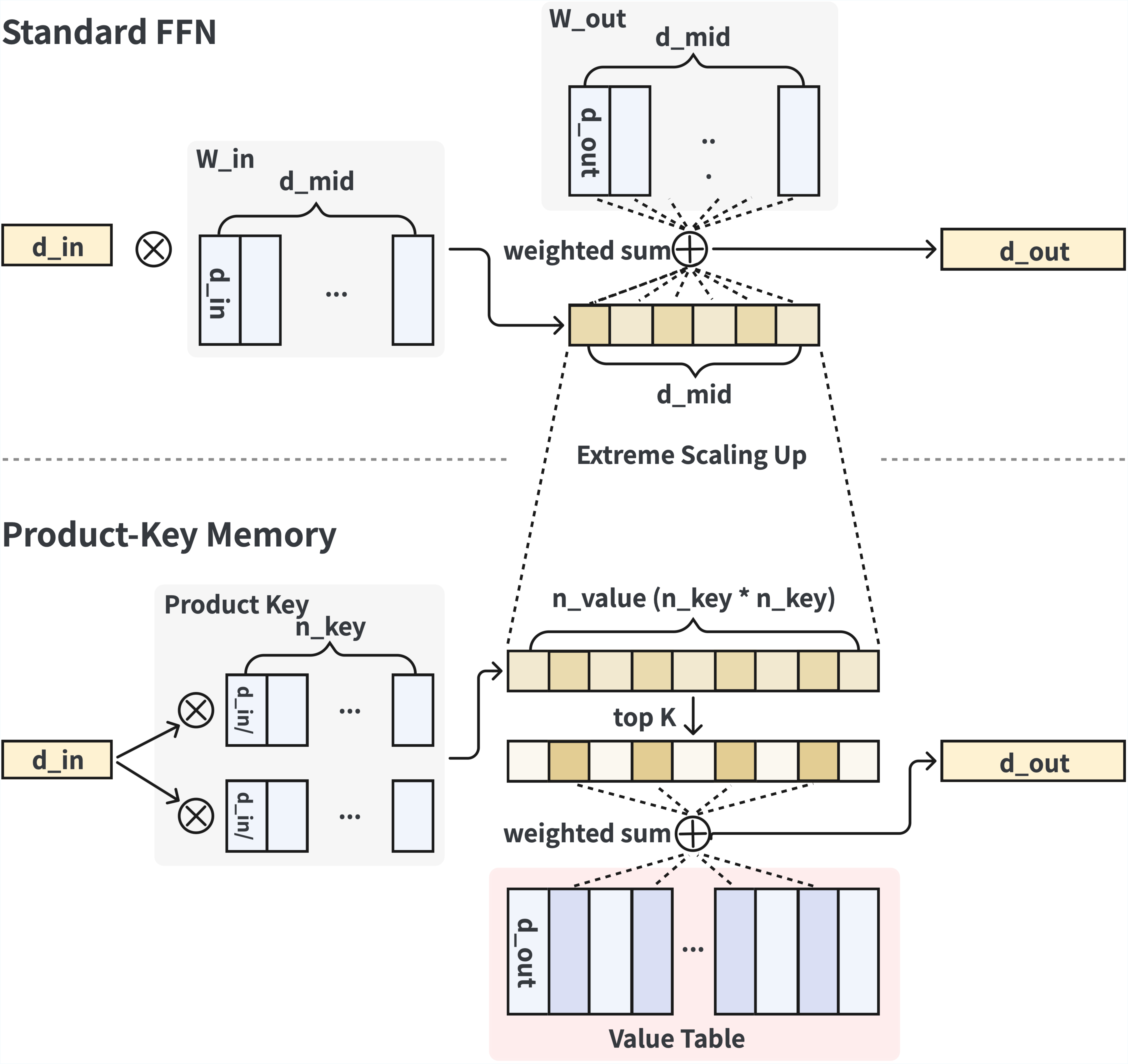}
\vspace{-5pt}
\caption{Comparison with stardard FFN and MSN.}
\label{fig:ffn_vs_mem}
\vspace{-5pt}
\end{figure}

In this part, we further compare MSN with feed-forward network (FFN) and SMoE from a structural perspective. Specifically, a standard FFN and SMoE can be formulated as follows.

\begin{equation}\label{eq:2LFFN}
\text{(standard FFN:)}\quad\bm{v}_o^{\text{FFN}} = \phi(\bm{x}_{\text{in}}\bm{W}_{\text{in}}+\bm{b_1})\bm{W}_{\text{out}} + \bm{b_2} 
                  = \sum_{i=1}^{d_{\text{mid}}} h_i \bm{W}_{\text{out}(i,:)} + \bm{b_2},
\end{equation}
where $\bm{x}_{\text{in}}$ is the input vector, $\phi(\cdot)$ is the non-linear activation function, and $\bm{h} \in \mathbb{R}^{d_{\text{mid}}}$ represents the intermediate hidden state. The weight matrices $\bm{W}_{\text{in}} \in \mathbb{R}^{d_{\text{in}} \times d_{\text{mid}}}$ and $\bm{W}_{\text{out}} \in \mathbb{R}^{d_{\text{mid}} \times d_{\text{out}}}$ correspond to the expansion and projection layers, respectively. $\bm{v}_o^{\text{FFN}} \in \mathbb{R}^{d_\text{out}}$ denotes the final output vector, formulated as a weighted aggregation of the projection weights conditioned on hidden unit activations
\begin{equation}\label{eq:SMoE}
\text{(SMoE:)}\quad\bm{v}_o^{\text{SMoE}}=\sum_{i\in\mathcal{G}}g_i \, \text{FFN}_i(\bm{x}_{\text{in}}),
\end{equation}
where $\mathcal{G}$ denotes the set of active experts, and $g_i$ are the corresponding gating weights.

By comparing Equations~\eqref{eq:2LFFN} and \eqref{eq:vo_mem}, we observe that unlike standard FFN, MSN's intermediate activations $S[i,j]$ derive primarily from memory values $\mathcal{V}$, whose parameter scale substantially exceeds that of $\bm{W}_{\text{in/out}}$ (Figure~\ref{fig:ffn_vs_mem}). 
As a result, MSN's output $\bm{v}_o$ can be viewed as a \emph{sparse approximation} of a standard FFN, while its theoretical modeling capacity exceeds conventional FFN due to the expansive memory space. While the output of SMoE in Equation~\eqref{eq:SMoE} can also be viewed as a sparse combination of FFNs, its capacity for sparsity is inherently limited. This is because each FFN expert exhibits quadratic scaling in parameters with respect to the hidden size, constraining the overall sparsification capability. In contrast, the design of MSN enables more effective model capacity scaling under the same computational constraints.

\subsection{Stabilized and Balanced Optimization}
\label{subsec:stability}

\subsubsection{Avoiding Extreme Similarity Distributions}

In practice, we observe that the high variance in PKM similarity scores often leads to peaked softmax distributions, causing gradient concentration and potential key-value collapse. To ensure training stability and load balance, we adopt the two following strategies:

\begin{itemize}[leftmargin=*]
\item \textbf{Layer Normalization.} We apply Layer Normalization~\cite{ba2016layer} to both query and sub-key vectors prior to similarity computation. This effectively stabilizes the score distribution and thus reduces the risk of softmax saturation.
\item \textbf{Learning Rate Warm-up.} We linearly warm up the learning rate from 0.1\% to the base value over the first 50k steps. This initial low-gradient phase encourages diverse value exploration, ensuring load balance and sufficient training for all value slots.
\end{itemize}

\subsubsection{Achieving Representation Update via Over-Parameterization}
As shown in Equation~\eqref{eq:vo_mem}, under extreme load imbalance where the activated memory slots are dominated by a very small subset, the memory network approximately degenerates into a dense network with a hidden size $k$ (\ie the activation count). Consequently, load imbalance tends to have a more severe impact on MSN than on SMoE architectures. This not only reduces the representational capacity of the output $\bm{v}_o$, but also adversely affects the update of the sub-key parameters ($\mathcal{K}_{\text{row}}$, $\mathcal{K}_{\text{col}}$), ultimately leading to catastrophic memory embedding collapse. Specifically, a memory key representation $\bm{k}_i=\mathcal{K}_{\text{row}}[i]$ is updated as follows
\begin{equation*}
    \bm{k}_i^{(t+1)} = \bm{k}_i^{(t)} - \eta \frac{\partial \mathcal{L}}{\partial \bm{k}_i^{(t)}},
\end{equation*}
where $\eta$ denotes the learning rate and $t$ denotes the step index. Thus, unselected keys $\bm{k}_i$ in a batch receive no gradients, leaving their embeddings not updated.

To tackle this challenge, inspired by \emph{SimVQ}~\cite{zhu2025addressing}, we adopt a simple-yet-effective \emph{key-over-parameterization strategy}. Specifically, we introduce additional learnable linear transformations for the sub-keys as follows
\begin{equation*}
\tilde{\mathcal{K}}_{\text{row}} = \mathbf{\Theta}_{\text{row}} \mathcal{K}_{\text{row}}, \ 
\tilde{\mathcal{K}}_{\text{col}} = \mathbf{\Theta}_{\text{col}} \mathcal{K}_{\text{col}}.
\end{equation*}
Based on above, the subsequent similarity computations are performed using the transformed memory values instead of Equation~\eqref{eq:S_old} as follows
\begin{equation*}
S_{\text{row}} = \tilde{\mathcal{K}}_{\text{row}}\bm{q}_{\text{row}},\  S_{\text{col}} = \tilde{\mathcal{K}}_{\text{col}}\bm{q}_{\text{col}}.
\end{equation*}
To theoretically demonstrate the effectiveness of the above over-parameterization strategy, taking $\mathbf{\Theta}_{\text{row}}$ as example, we can compute the gradient propagation and parameter update as follows
\begin{equation*}
\begin{aligned}
\mathbf{\Theta}_{\text{row}}^{(t+1)} &= \mathbf{\Theta}_{\text{row}}^{(t)} - \eta \frac{\partial \mathcal{L}}{\partial \mathbf{\Theta}_{\text{row}}^{(t)}}
           = \mathbf{\Theta}_{\text{row}}^{(t)} - \eta \frac{\partial \mathcal{L}}{\partial \mathcal{K}^{(t)}} \,\mathcal{K}^{(t)\top}, \\
\mathcal{K}^{(t+1)} &= \mathcal{K}^{(t)} - \eta \frac{\partial \mathcal{L}}{\partial \mathcal{K}^{(t)}}
           = \mathcal{K}^{(t)} - \eta \mathbf{\Theta}_{\text{row}}^{(t)\top} \frac{\partial \mathcal{L}}{\partial \tilde{\mathcal{K}}^{(t)}}, \\
\tilde{\bm{k}}_i^{(t+1)} &= \mathbf{\Theta}_{\text{row}}^{(t+1)} \,\bm{k}_i^{(t+1)} \\ 
           &\approx \tilde{\bm{k}}_i^{(t)} - \eta \left( \frac{\partial \mathcal{L}}{\partial \mathcal{K}^{(t)}} \,\mathcal{K}^{(t)\top} \bm{k}_i^{(t)}
           + \mathbf{\Theta}_{\text{row}}^{(t)} \mathbf{\Theta}_{\text{row}}^{(t)\top} \frac{\partial \mathcal{L}}{\partial \tilde{\bm{k}}_i^{(t)}} \right).
\end{aligned}
\end{equation*}
From the above equations, we can observe that even when the similarity scores become highly concentrated, gradients can still propagate through the over-parameterized transformations, thereby enabling the over-parameterized memory vector to be updated and improving the training robustness. 

\subsection{Memory-Input Fusion via Gating}
\label{subsec:memory-gate}

After obtaining the memory output $\bm{v}_o$, a natural question is \emph{how to effectively integrate it into the existing model architecture to fully exploit the retrieved memory}. 
A straightforward approach is to directly use the retrieved memory representation $\bm{v}_o$ as input to subsequent layers, or fuse it with the original input $\bm{x}_{\text{in}}$ via concatenation. However, we observed only marginal improvements in practice. 
As indicated by Equation~\eqref{eq:vo_mem}, the number of sparsely activated memory slots is typically much smaller than the input's dimension (\ie $d\gg k$). 
Thus, we postulate that naively replacing $\bm{x}_{\text{in}}$ with $\bm{v}_o$ or treating $\bm{v}_o$ merely as incremental information fails to adequately preserve or enhance the original input's information, finally resulting in suboptimal performance.

To address this issue, we propose a \emph{Memory-Gated Fusion} strategy, where the memory output $\bm{v}_o$ is used as a sparse gating signal to modulate the input $\bm{x}_{\text{in}}$ in a element-wise manner as follows
\begin{equation}\label{eq:gating}
\tilde{\bm{x}} = \bm{x}_{\text{in}} \odot \tanh (\bm{v}_o),
\end{equation}
where $\tanh(x)=\frac{e^x-e^{-x}}{e^x+e^{-x}}$.
The gated representation $\tilde{\bm{x}}$ is then passed to subsequent modeling modules, such as FFN or SMoE blocks. In practice, we find that inserting the Memory-Gated module into only a few feature interaction blocks (typically $1$ or $2$) achieves the best trade-off between model capacity and inference efficiency. This design preserves original input information while incorporating dynamically activated sparse memory parameters, thus enhancing representation capacity and personalization performance without computational overhead.

\subsection{Training and Deployment Optimization}\label{subsec:deployment}

To address the bottlenecks in Top-$k$ retrieval and weighted aggregation within MSN, we utilize two specialized operators for enhancing both training and serving efficiency:

\begin{itemize}[leftmargin=*]
\item \textbf{Sparse-Gather Op} fuses sparse access, weight computation, and gradient updates into a unified kernel. To overcome I/O bounds, we optimize task partitioning and vectorized memory access to fully saturate bandwidth. This integration minimizes redundant memory I/O, achieving near-peak performance across various accelerator architectures.
\item \textbf{AirTopK Op.} The native TensorFlow Top-$k$ operator uses radix sort, which is suboptimal for our specific retrieval scale (32 out of 1024 candidates). We instead adopt \emph{AirTopK}~\cite{zhang2023parallel}, leveraging shared-memory-based fast sorting to minimize retrieval latency. Consequently, the average and P99 latency overheads are reduced from 13ms and 26ms, respectively, to levels comparable with the production baseline.
\end{itemize}

\begin{tcolorbox}[colback=gray!7,colframe=black!30,boxrule=0.8pt,arc=1mm]
\textbf{In summary}, MSN provides three key advantages:
\begin{itemize}[leftmargin=*]
\item \textbf{Efficient Scaling.} By combining large-scale memory value storage with sub-key factorization, MSN greatly expands parameter space while maintaining sublinear computational complexity relative to scale.

\item \textbf{Adaptive Personalization.} The memory-gated design acts as a generalized adaptive parameterization (\eg LHUC~\cite{lhuc} and Adaptive Domain Scaling~\cite{chai2025adaptive}), leveraging a vast parameter space to generate input-specific modulation for fine-grained personalization.

\item \textbf{Unified Framework.} MSN can be seamlessly integrated into different feature interaction modules, including FFN- and SMoE-based blocks. This unified design allows heterogeneous modeling components to share a common sparse activation scaling paradigm.
\end{itemize}
\end{tcolorbox}

\section{Experiments}
In this section, we first set up the experiments, and then present the results with detailed analysis.
\subsection{Experimental Setup}

\subsubsection{Datasets and Environment} The offline experiments are conducted on training data from the Douyin search system, which are collected from online interaction logs and user feedback signals. The dataset contains over 1,000 heterogeneous features, including numerical features, categorical ID features, cross features, and sequential behavioral features. These features cover hundreds of millions of users and billions of videos. All ID-related features are anonymized before training to ensure user privacy and data security. The sparse input features are then encoded as dense embeddings via embedding lookup, which serve as inputs to the recommendation model. The dataset includes billions of user–item interaction records per day. For all experiments, we use data collected over a continuous three-week period for training and evaluation.

Following previous works~\cite{han2025lemur,huang2026hyformer,MDL}, the CTR prediction model is cold-started for offline evaluation and warmed up with checkpoints for online evaluation. All baselines follow the same batch size and optimizer settings. All experiments are conducted on 96-GPUs in a hybrid distributed training framework.

\subsubsection{Evaluation Metrics} For offline evaluation, we use QAUC (Query-Level AUC) as the primary metrics. QAUC computes AUC (Area Under the Curve) within each query and then averages the scores over all queries, which is commonly regarded as more aligned with online performance in industrial search scenarios. In addition, we report the total number of dense parameters, the number of activated parameters together with the activation ratio, and the training FLOPs measured with a batch size of 2048.   

\subsubsection{Implementation Details}

\begin{figure}[t]
\centering
\includegraphics[width=\columnwidth]{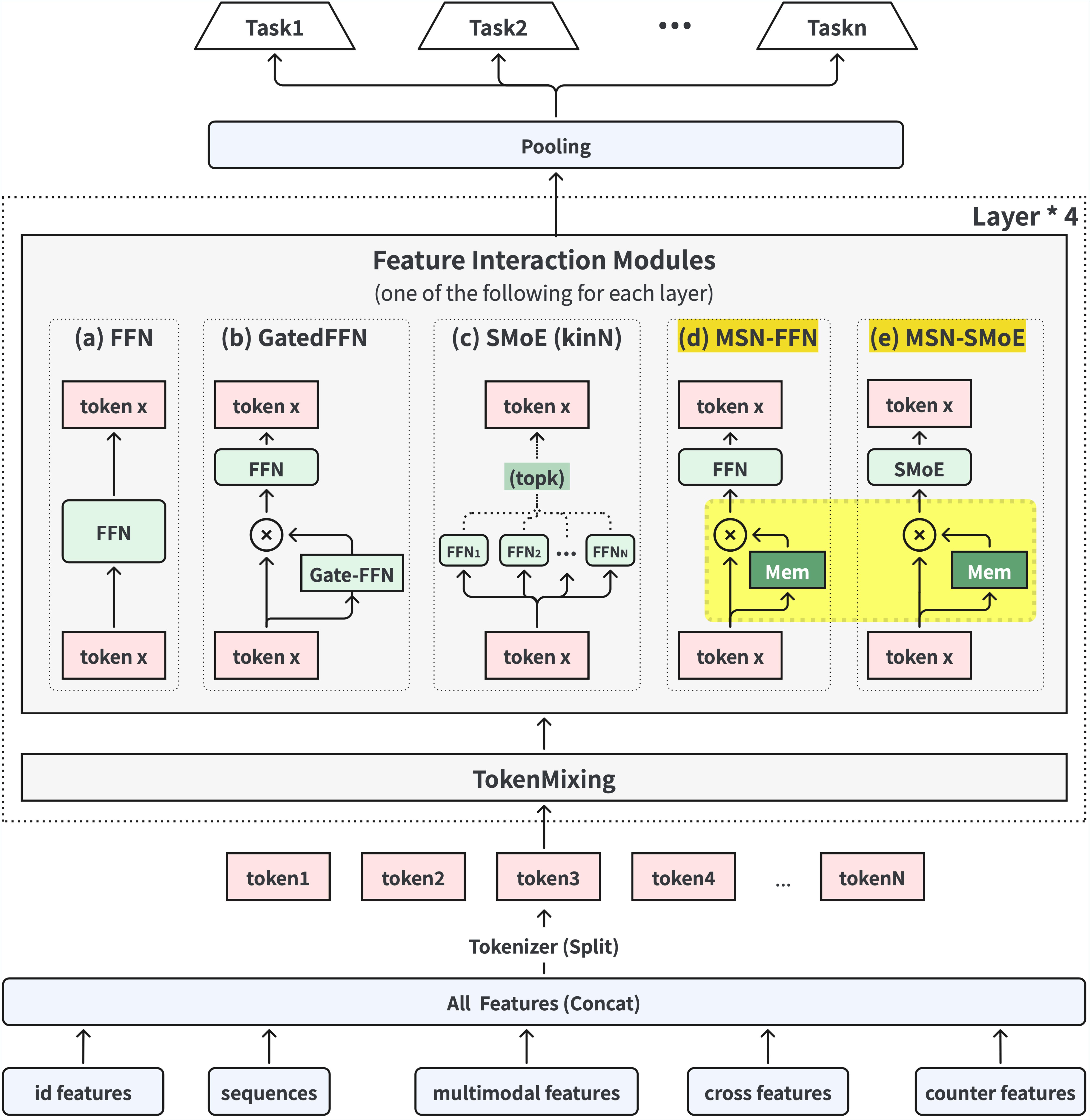}
\vspace{-8pt}
\caption{The overall architecture of the backbone model.}
\label{fig:exp}
\vspace{-5pt}
\end{figure}

In our experiments, the backbone model is the latest version of the Personalized Ranking Model used in Douyin Search, which is built upon a RankMixer-based architecture~\cite{zhu2025rankmixer} in Figure~\ref{fig:exp}. 
We also compare the backbone model with DLRM-MLP (adopting the vanilla MLP for feature crossing) and Wukong~\cite{zhang2024wukong} (adopting DHEN with Factorization Machine Block and Linear Compress Block for model scaling). 

To further investigate the scaling behavior of MSN in comparison with traditional methods, we replace the feature interaction scaling modules in different backbone layers and adjust their hidden sizes, so that all models have comparable parameter counts. When integrating MSN into the backbone, we insert the MSN module before pertoken FFN or SMoE layers, referred to as MSN-FFN and MSN-SMoE, respectively. We implement a hybrid parameter allocation strategy: within each layer, all tokens access a shared set of memory values $\mathcal{V}$ ($\mathcal{V}$ differs across layers); conversely, the query networks and memory key parameters ($\mathcal{K}_{\text{row}}$ and $\mathcal{K}_{\text{col}}$) remain token-specific and independent. Unless otherwise specified, we set the default memory size to $n=256^2$, retrieval size to $k=32$ and the number of MSN-equipped layers to $L=2$.

\label{sec:exp}

\subsection{Overall Performance}

\begin{table}[t]
\centering
\caption{Overall Performance \& Efficiency Comparison.}
\vspace{-8pt}
\label{tab:overall}
\renewcommand\arraystretch{1.2}
\setlength{\tabcolsep}{1.6mm}{
\resizebox{0.9\columnwidth}{!}{
\begin{tabular}{lcccc}
\toprule
\multicolumn{1}{c}{\multirow{2}{*}{\textbf{Methods}}} & \multicolumn{2}{c}{\textbf{QAUC$\uparrow$}} & \multicolumn{2}{c}{\textbf{Efficiency}}   \\ \cmidrule{2-5}

\multicolumn{1}{c}{}  & \textbf{Click}    & \textbf{Finish}   & \textbf{\#Params} & \textbf{FLOPs\small{/Batch}} \\
\midrule
DLRM-MLP   & 0.6731 & 0.7545 & 118M/118M & 520G \\
Wukong &  +0.25\% & +0.13\% & 127M/127M & 567G \\
\midrule
\multicolumn{5}{c}{\small{The modules below are based on \textbf{RankMixer} in Figure~\ref{fig:exp}.}} \\
\midrule
FFN   & +0.35\% & +0.21\% & 120M/120M & 529G \\
GatedFFN & +0.35\% & +0.21\% & 122M/122M & 541G \\
SMoE (2in4) & +0.38\% & +0.22\% & 120M/194M & 543G \\
SMoE (2in6) & +0.43\% & +0.25\% & 120M/267M & 546G \\
\midrule
\textbf{MSN-FFN} & \textbf{+0.54\%} & \textbf{+0.33\%} & 122M/252M & 556G \\
\textbf{MSN-SMoE} & \textbf{+0.50\%} & \textbf{+0.31\%} & 122M/294M & 560G \\
\bottomrule
\end{tabular}}}
\vspace{-5pt}
\end{table}

The overall performance comparison of baseline methods and MSN is shown in Table~\ref{tab:overall}. We select Click and Finish as the primary objects, and measure the QAUC improvement over the baseline to evaluate model effectiveness. Since both SMoE and MSN adopt sparse scaling techniques, we report the number of activated parameters during inference $p_{\text{act}}$ and the total number of model parameters $p_{\text{tot}}$ in the \#Params column, formatted as $p_{\text{act}}/p_{\text{tot}}$. Our key observations are listed as follows:

\begin{itemize}[leftmargin=*] 
\item Under a comparable activation budget of 120M parameters, models built on the RankMixer backbone achieve the strongest performance across all metrics, demonstrating the effectiveness of its feature interaction capability. 
\item Within the RankMixer backbone, under the same parameter budget, Gated-FFN does not provide noticeable improvements over standard FFN, suggesting that architectural gating alone is insufficient for further performance gains. 
\item We then compare MSN with SMoE, where each token selects Top‑$k$ from $N$ independent experts. We test two configurations: 120M activated out of 190M total parameters (2‑out‑of‑4 experts, denoted as 2in4) and 120M activation out of 267M total (2in6). Although increasing the total parameter size via SMoE leads to performance improvements, its results remain inferior to MSN under a comparable activated capacity. In contrast, our MSN replaces part of the original layers with memory-gated modules, introducing memory-based sparse parameters, and achieves the highest QAUC among all compared methods.
\item We further implement two MSN variants, MSN-FFN and MSN-SMoE, by adapting the framework to both dense FFN and SMoE backbones. Both variants consistently outperform their corresponding baselines (FFN and SMoE), demonstrating the general effectiveness of the proposed framework. Counter-intuitively, although MSN-SMoE introduces a larger total parameter size than MSN-FFN, it exhibits slightly inferior performance. A plausible explanation is that stacking sparse scaling mechanisms increases optimization difficulty, potentially hindering model convergence. This observation suggests that a straightforward gating-based combination of two distinct sparse modules may be sub-optimal, highlighting the need for more sophisticated integration strategies in future research. 
\end{itemize}

\subsection{In-Depth Analysis}

\begin{table}[t]
\centering
\caption{In-Depth Analysis on MSN Components. \textbf{Default MSN-FFN Configuration: }1 Layer with MSN \& Memory Value Size $n=256^2$ \& Retrieval Size $k=32$ \& Gating Function Tanh.}
\vspace{-8pt}
\label{tab:ablation}
\renewcommand\arraystretch{1.2}
\setlength{\tabcolsep}{1mm}{
\resizebox{\columnwidth}{!}{
\begin{tabular}{ccccc}
\toprule
\textbf{Ablation Topics} & \textbf{MSN Variants} & \textbf{$\Delta\text{QAUC}$} & \textbf{\#Params} & \textbf{MSN Param Ratio} \\
\midrule
\textbf{Default MSN-FFN} & -- & -- & 122M/194M & 37.1\% \\
\midrule
\multirow{2}{*}{\textbf{\#Layer with MSN}}
   & $L$=2 & +0.05\% & 122M/252M & 51.6\% \\
   & $L$=4 & -0.05\% & 122M/327M & 62.9\% \\
\midrule
\multirow{2}{*}{\textbf{Scale of MSN Param}}  & n=$512^2$ & +0.04\% & 122M/398M & 69.3\% \\
   & pertoken$\mathcal{V}$ & +0.06\% & 122M/1.19B & 89.7\% \\
\midrule
\multirow{3}{*}{\textbf{Top-$k$ Retrieval Size}} & $k$=\text{64} & +0\% & \multirow{3}{*}{122M/194M}   & \multirow{3}{*}{37.1\%} \\
   & $k$=\text{16} & -0.01\% & & \\
   & $k$=\text{8} & -0.05\% & & \\
\midrule
\multirow{2}{*}{\textbf{MSN Gating Function}}  & w/o Tanh & -0.02\% & \multirow{2}{*}{122M/194M} & \multirow{2}{*}{37.1\%} \\
   & Sigmoid & -0.05\% & & \\
\bottomrule
\end{tabular}}}
\vspace{-5pt}
\end{table}

To explore the optimal architectural design of MSN, we conduct a set of ablation studies to study the impact of four key factors: the number of layers equipped with MSN, the scale of sparse parameters, the retrieval top‑$k$ size, and the memory-input fusion's gating function. The results are summarized in Table~\ref{tab:ablation}, and our analysis for each variable is presented below.

\subsubsection{Number of Layers Equipped with MSN}

When adopting MSN as the scaling module, we replace $x$ out of the four backbone layers with MSN-FFN, while the remaining $4-x$ layers use the default FFN, denoted as MSN-$x$L. With varying number of layers in which MSN is applied, the results and analyses are summarized as follows:

\begin{itemize}[leftmargin=*]
    \item When inserting MSN into one single layer (the default MSN-FFN or MSN-1L), the total number of parameters increases to 194M, with memory-related parameters accounting for 37.1\%. 
    \item Extending MSN to two layers (MSN-2L) further increases the total parameter count to 252M, with memory parameters comprising 51.6\% of the model. In this setting, QAUC improves by +0.05\%, substantially outperforming the single-layer configuration. This suggests that moderately introducing MSN across multiple layers can further strengthen personalization and lead to better performance.
    \item However, when MSN is applied to all four layers (MSN-4L), although the total parameter size continues to grow to 327M with 62.9\% memory parameters, the QAUC gain drops to -0.05\%, underperforming both the one-layer and two-layer configurations. We conjecture that this is caused by excessive sparsification, which limits effective feature interaction and harms model learning.
\end{itemize}

\subsubsection{Scale of Sparse Parameters}

While keeping the activated parameter size fixed at 122M, we expand the sparse parameter space of MSN in two ways and observe consistent performance gains:
\begin{itemize}[leftmargin=*]
    \item \textbf{Enlarging the memory value space} (MSN‑$n=512^2$): Increasing $ n $ from $ 256^2 $ to $ 512^2 $ raises the total parameters to 398M (with memory parameters constituting 69.3\%). This leads to a QAUC improvement of +0.04\%.
    \item \textbf{Adopting pertoken memory values} (pertoken$\mathcal{V}$): While keeping $ n = 256^2 $, we assign independent memory values to each token instead of sharing them across tokens. This expands the sparse parameters to 1.19B (89.7\% memory share) and yields a QAUC gain of +0.06\%, the best among all tested configurations.
\end{itemize}
These results demonstrate that, under a fixed activated‑parameter budget, enlarging the sparse parameter space of the memory network effectively enhances the model's ability to store and retrieve fine‑grained user behavior patterns, thereby improving the CTR prediction accuracy.

\subsubsection{Top-$k$ Retrieval Size}

We conduct an ablation study on the retrieval size of memory values to examine its impact on CTR prediction performance. The results can be summarized as follows.
\begin{itemize}[leftmargin=*]
\item When $k\geq 32$, the model exhibits stable and strong performance. Specifically, reducing the Top-$k$ value from 64 to 32 results in almost no change in QAUC. This indicates that, when sparse parameters are sufficiently utilized, the majority of useful information is concentrated within selected Top-$k$ memory entries. Increasing the Top-$k$ size further introduces limited additional information and yields marginal performance gains.
\item However, when $k$ is reduced to 8, the model performance degrades sharply, with QAUC improvement dropping -0.05\%, which is notably worse than configurations with $k\geq 32$. This degradation is primarily caused by insufficient utilization of sparse memory values: since $k$ determines the number of slots retrieved from the sparse memory space, a too-small Top-$k$ not only substantially limits the activation of sparse parameters, leaving a large portion of the memory underutilized, but also truncates relevant information that lies beyond the selected entries. These results further highlight that maintaining a sufficient activation scale of sparse memory parameters is a critical prerequisite for achieving strong model performance.
\end{itemize}

\subsubsection{Selection of Memory Gating Function}
We conduct an ablation study on the gating function in Equation~\eqref{eq:gating}, replacing the default tanh with a sigmoid activation or an identity function (i.e., removing the activation entirely). As shown in the last two rows of Table~\ref{tab:ablation}, both alternative gating choices lead to a clear drop in recommendation performance compared to the default MSN‑FFN. These results confirm that the tanh-based gating is an effective component of the proposed Memory‑Gated Fusion strategy.

\begin{figure}[t]
\centering
\includegraphics[width=\columnwidth]{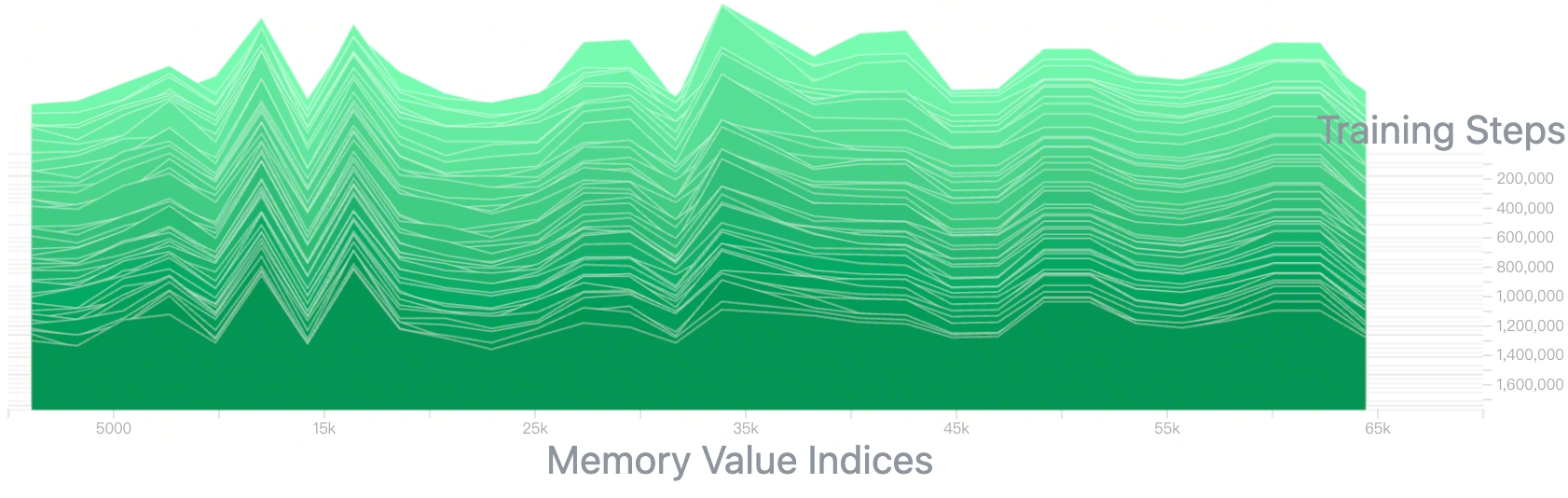}
\vspace{-10pt}
\caption{The distribution of activated memory values.}
\label{fig:mem_activate}
\vspace{-5pt}
\end{figure}

\subsubsection{Distribution of Activated Memory Values}

To examine whether each memory value is utilized effectively during training, we visualize the activation count of every memory slot in Figure~\ref{fig:mem_activate}. The results show that all memory values are sufficiently and frequently activated, with no significant imbalance in the activation distribution. This indicates that the entire memory table $\mathcal{V}$ receives adequate gradient updates during training, which in turn enhances MSN's ability to improve recommendation performance.

\subsection{Online A/B Tests}

This section reports the online performance of MSN in Douyin Search application. The baseline is a well‑optimized RankMixer-based model that has served hundreds of millions of users for several months. For online evaluation, we focus on key user engagement metrics: \emph{Active Day} (the average number of active days per user, reflecting user retention in search), \emph{Query Change Rate}, \emph{Watch Time}, \emph{Video Finish Rate}, as these metrics reflect whether user search intent is well satisfied. 
Specifically, \emph{Query Change Rate} measures the probability of users manually refining a search query into a more specific one (\eg changing from "dog food" to "doggie's foods"). It is calculated as follows
\begin{equation*}
    \text{Query Change Rate} = 
    \frac{
\text{\# user-query pairs with query refining}
}{
\text{\# distinct user-query pairs}
}.
\end{equation*}

As shown in Table~\ref{tab:abtest}, MSN consistently outperforms the baseline. On the overall traffic, MSN achieves consistent improvements across key engagement metrics: Active Days increase by 0.0503\%, while the Query Change Rate decreases by 0.1337\%. Furthermore, consumption metrics show significant uplift, with Average Watch Time per User and Finish Rate rising by 0.2958\% and 0.2071\%, respectively. Moreover, in Double Column search scenario, which is particularly critical for user engagement in Douyin Search, MSN maintains its superiority, yielding a 0.0481\% gain in Active Days and a 0.275\% reduction in Query Change Rate. These results demonstrates that MSN delivers significant real‑world benefits when deployed at scale in an industrial recommender system.

\begin{table}[t]
\centering
\caption{Online A/B Test Results on Douyin Search.}
\vspace{-8pt}
\label{tab:abtest}
\renewcommand\arraystretch{1.2}
\setlength{\tabcolsep}{1.2mm}{
\resizebox{0.99\columnwidth}{!}{
\begin{tabular}{ccccc}
\toprule
\textbf{Metrics} &  Active Days$\uparrow$ &Change Query$\downarrow$ & Watch Time$\uparrow$ & Finish Play$\uparrow$ \\
\midrule
\textbf{Overall} &  +0.0503\% &-0.1337\% & +0.2958\% & +0.2071\% \\
\textbf{Double Column} & +0.0481\% & -0.275\% & +0.3032\% & +0.4549\% \\
\bottomrule
\end{tabular}}}
\end{table}

\section{Conclusion}

In this paper, we presented MSN, a memory-based sparse scaling framework for improving the capacity of DLRMs under strict efficiency constraints. Different from existing scaling approaches that rely on enlarging embedding tables, feature interaction modules, or activating a small number of large experts, MSN scales model capacity through lightweight and efficient parameter retrieval from a large key-value memory. By combining Product-Key Memory for efficient retrieval, a Memory Gating Mechanism for seamless integration with downstream feature interaction modules, and a set of normalization and over-parameterization strategies for stable optimization, MSN achieves strong personalization capability with controlled computational and memory access cost. Extensive experiments show that MSN consistently improves recommendation performance across different model backbones. More importantly, the framework has been successfully deployed in the Douyin Search Ranking System, bringing clear gains in both offline evaluation and online metrics, which demonstrates its practicality in large-scale industrial environments. In the future, we plan to explore how memory-based scaling can be combined with other forms of sparsity, such as sparse feature interaction or dynamic computation depth, to further improve efficiency.

\newpage
\bibliographystyle{ACM-Reference-Format}
\balance
\bibliography{sample-base}


\begin{thebibliography}{49}


\ifx \showCODEN    \undefined \def \showCODEN     #1{\unskip}     \fi
\ifx \showISBNx    \undefined \def \showISBNx     #1{\unskip}     \fi
\ifx \showISBNxiii \undefined \def \showISBNxiii  #1{\unskip}     \fi
\ifx \showISSN     \undefined \def \showISSN      #1{\unskip}     \fi
\ifx \showLCCN     \undefined \def \showLCCN      #1{\unskip}     \fi
\ifx \shownote     \undefined \def \shownote      #1{#1}          \fi
\ifx \showarticletitle \undefined \def \showarticletitle #1{#1}   \fi
\ifx \showURL      \undefined \def \showURL       {\relax}        \fi
\providecommand\bibfield[2]{#2}
\providecommand\bibinfo[2]{#2}
\providecommand\natexlab[1]{#1}
\providecommand\showeprint[2][]{arXiv:#2}

\bibitem[Ba et~al\mbox{.}(2016)]%
        {ba2016layer}
\bibfield{author}{\bibinfo{person}{Jimmy~Lei Ba}, \bibinfo{person}{Jamie~Ryan Kiros}, {and} \bibinfo{person}{Geoffrey~E Hinton}.} \bibinfo{year}{2016}\natexlab{}.
\newblock \showarticletitle{Layer normalization}.
\newblock \bibinfo{journal}{\emph{arXiv preprint arXiv:1607.06450}} (\bibinfo{year}{2016}).
\newblock


\bibitem[Berges et~al\mbox{.}(2025)]%
        {bergesmemory}
\bibfield{author}{\bibinfo{person}{Vincent-Pierre Berges}, \bibinfo{person}{Barlas Oguz}, \bibinfo{person}{Daniel HAZIZA}, \bibinfo{person}{Wen-tau Yih}, \bibinfo{person}{Luke Zettlemoyer}, {and} \bibinfo{person}{Gargi Ghosh}.} \bibinfo{year}{2025}\natexlab{}.
\newblock \showarticletitle{Memory Layers at Scale}. In \bibinfo{booktitle}{\emph{ICML}}.
\newblock


\bibitem[Blondel et~al\mbox{.}(2016)]%
        {hofm}
\bibfield{author}{\bibinfo{person}{Mathieu Blondel}, \bibinfo{person}{Akinori Fujino}, \bibinfo{person}{Naonori Ueda}, {and} \bibinfo{person}{Masakazu Ishihata}.} \bibinfo{year}{2016}\natexlab{}.
\newblock \showarticletitle{Higher-order factorization machines}.
\newblock \bibinfo{journal}{\emph{NeurIPS}}  \bibinfo{volume}{29} (\bibinfo{year}{2016}).
\newblock


\bibitem[Chai et~al\mbox{.}(2025)]%
        {chai2025adaptive}
\bibfield{author}{\bibinfo{person}{Zheng Chai}, \bibinfo{person}{Hui Lu}, \bibinfo{person}{Di Chen}, \bibinfo{person}{Qin Ren}, \bibinfo{person}{Yuchao Zheng}, {and} \bibinfo{person}{Xun Zhou}.} \bibinfo{year}{2025}\natexlab{}.
\newblock \showarticletitle{Adaptive Domain Scaling for Personalized Sequential Modeling in Recommenders}. In \bibinfo{booktitle}{\emph{SIGIR}}. \bibinfo{pages}{4234--4238}.
\newblock


\bibitem[Cheng et~al\mbox{.}(2016)]%
        {widedeep}
\bibfield{author}{\bibinfo{person}{Heng-Tze Cheng}, \bibinfo{person}{Levent Koc}, \bibinfo{person}{Jeremiah Harmsen}, \bibinfo{person}{Tal Shaked}, \bibinfo{person}{Tushar Chandra}, \bibinfo{person}{Hrishi Aradhye}, \bibinfo{person}{Glen Anderson}, \bibinfo{person}{Greg Corrado}, \bibinfo{person}{Wei Chai}, \bibinfo{person}{Mustafa Ispir}, {et~al\mbox{.}}} \bibinfo{year}{2016}\natexlab{}.
\newblock \showarticletitle{Wide \& deep learning for recommender systems}. In \bibinfo{booktitle}{\emph{Proceedings of the 1st workshop on deep learning for recommender systems}}. \bibinfo{pages}{7--10}.
\newblock


\bibitem[Cheng et~al\mbox{.}(2026)]%
        {cheng2026conditional}
\bibfield{author}{\bibinfo{person}{Xin Cheng}, \bibinfo{person}{Wangding Zeng}, \bibinfo{person}{Damai Dai}, \bibinfo{person}{Qinyu Chen}, \bibinfo{person}{Bingxuan Wang}, \bibinfo{person}{Zhenda Xie}, \bibinfo{person}{Kezhao Huang}, \bibinfo{person}{Xingkai Yu}, \bibinfo{person}{Zhewen Hao}, \bibinfo{person}{Yukun Li}, {et~al\mbox{.}}} \bibinfo{year}{2026}\natexlab{}.
\newblock \showarticletitle{Conditional Memory via Scalable Lookup: A New Axis of Sparsity for Large Language Models}.
\newblock \bibinfo{journal}{\emph{arXiv preprint arXiv:2601.07372}} (\bibinfo{year}{2026}).
\newblock


\bibitem[Covington et~al\mbox{.}(2016)]%
        {covington2016deep}
\bibfield{author}{\bibinfo{person}{Paul Covington}, \bibinfo{person}{Jay Adams}, {and} \bibinfo{person}{Emre Sargin}.} \bibinfo{year}{2016}\natexlab{}.
\newblock \showarticletitle{Deep neural networks for youtube recommendations}. In \bibinfo{booktitle}{\emph{RecSys}}. \bibinfo{pages}{191--198}.
\newblock


\bibitem[Gui et~al\mbox{.}(2023)]%
        {gui2023hiformer}
\bibfield{author}{\bibinfo{person}{Huan Gui}, \bibinfo{person}{Ruoxi Wang}, \bibinfo{person}{Ke Yin}, \bibinfo{person}{Long Jin}, \bibinfo{person}{Maciej Kula}, \bibinfo{person}{Taibai Xu}, \bibinfo{person}{Lichan Hong}, {and} \bibinfo{person}{Ed~H Chi}.} \bibinfo{year}{2023}\natexlab{}.
\newblock \showarticletitle{Hiformer: Heterogeneous feature interactions learning with transformers for recommender systems}.
\newblock \bibinfo{journal}{\emph{arXiv preprint arXiv:2311.05884}} (\bibinfo{year}{2023}).
\newblock


\bibitem[Guo et~al\mbox{.}(2017)]%
        {guo2017deepfm}
\bibfield{author}{\bibinfo{person}{Huifeng Guo}, \bibinfo{person}{Ruiming Tang}, \bibinfo{person}{Yunming Ye}, \bibinfo{person}{Zhenguo Li}, {and} \bibinfo{person}{Xiuqiang He}.} \bibinfo{year}{2017}\natexlab{}.
\newblock \showarticletitle{DeepFM: a factorization-machine based neural network for CTR prediction}. In \bibinfo{booktitle}{\emph{IJCAI}}. \bibinfo{pages}{1725--1731}.
\newblock


\bibitem[Guo et~al\mbox{.}(2024)]%
        {guo2024embedding}
\bibfield{author}{\bibinfo{person}{Xingzhuo Guo}, \bibinfo{person}{Junwei Pan}, \bibinfo{person}{Ximei Wang}, \bibinfo{person}{Baixu Chen}, \bibinfo{person}{Jie Jiang}, {and} \bibinfo{person}{Mingsheng Long}.} \bibinfo{year}{2024}\natexlab{}.
\newblock \showarticletitle{On the Embedding Collapse when Scaling up Recommendation Models}. In \bibinfo{booktitle}{\emph{ICML}}. PMLR, \bibinfo{pages}{16891--16909}.
\newblock


\bibitem[Han et~al\mbox{.}(2025b)]%
        {han2025mtgr}
\bibfield{author}{\bibinfo{person}{Ruidong Han}, \bibinfo{person}{Bin Yin}, \bibinfo{person}{Shangyu Chen}, \bibinfo{person}{He Jiang}, \bibinfo{person}{Fei Jiang}, \bibinfo{person}{Xiang Li}, \bibinfo{person}{Chi Ma}, \bibinfo{person}{Mincong Huang}, \bibinfo{person}{Xiaoguang Li}, \bibinfo{person}{Chunzhen Jing}, {et~al\mbox{.}}} \bibinfo{year}{2025}\natexlab{b}.
\newblock \showarticletitle{Mtgr: Industrial-scale generative recommendation framework in meituan}. In \bibinfo{booktitle}{\emph{CIKM}}. \bibinfo{pages}{5731--5738}.
\newblock


\bibitem[Han et~al\mbox{.}(2025a)]%
        {han2025lemur}
\bibfield{author}{\bibinfo{person}{Xintian Han}, \bibinfo{person}{Honggang Chen}, \bibinfo{person}{Quan Lin}, \bibinfo{person}{Jingyue Gao}, \bibinfo{person}{Xiangyuan Ren}, \bibinfo{person}{Lifei Zhu}, \bibinfo{person}{Zhisheng Ye}, \bibinfo{person}{Shikang Wu}, \bibinfo{person}{XiongHang Xie}, \bibinfo{person}{Xiaochu Gan}, {et~al\mbox{.}}} \bibinfo{year}{2025}\natexlab{a}.
\newblock \showarticletitle{LEMUR: Large scale End-to-end MUltimodal Recommendation}.
\newblock \bibinfo{journal}{\emph{arXiv preprint arXiv:2511.10962}} (\bibinfo{year}{2025}).
\newblock


\bibitem[He and Chua(2017)]%
        {nfm}
\bibfield{author}{\bibinfo{person}{Xiangnan He} {and} \bibinfo{person}{Tat-Seng Chua}.} \bibinfo{year}{2017}\natexlab{}.
\newblock \showarticletitle{Neural factorization machines for sparse predictive analytics}. In \bibinfo{booktitle}{\emph{SIGIR}}. \bibinfo{pages}{355--364}.
\newblock


\bibitem[He(2024)]%
        {he2024mixture}
\bibfield{author}{\bibinfo{person}{Xu~Owen He}.} \bibinfo{year}{2024}\natexlab{}.
\newblock \showarticletitle{Mixture of a million experts}.
\newblock \bibinfo{journal}{\emph{arXiv preprint arXiv:2407.04153}} (\bibinfo{year}{2024}).
\newblock


\bibitem[Huang et~al\mbox{.}(2026b)]%
        {huang2026hyformer}
\bibfield{author}{\bibinfo{person}{Yunwen Huang}, \bibinfo{person}{Shiyong Hong}, \bibinfo{person}{Xijun Xiao}, \bibinfo{person}{Jinqiu Jin}, \bibinfo{person}{Xuanyuan Luo}, \bibinfo{person}{Zhe Wang}, \bibinfo{person}{Zheng Chai}, \bibinfo{person}{Shikang Wu}, \bibinfo{person}{Yuchao Zheng}, {and} \bibinfo{person}{Jingjian Lin}.} \bibinfo{year}{2026}\natexlab{b}.
\newblock \showarticletitle{HyFormer: Revisiting the Roles of Sequence Modeling and Feature Interaction in CTR Prediction}. In \bibinfo{booktitle}{\emph{SIGIR}}.
\newblock


\bibitem[Huang et~al\mbox{.}(2026a)]%
        {huang2025ultramemv2}
\bibfield{author}{\bibinfo{person}{Zihao Huang}, \bibinfo{person}{Yu Bao}, \bibinfo{person}{Qiyang Min}, \bibinfo{person}{Siyan Chen}, \bibinfo{person}{Ran Guo}, \bibinfo{person}{Hongzhi Huang}, \bibinfo{person}{Defa Zhu}, \bibinfo{person}{Yutao Zeng}, \bibinfo{person}{Banggu Wu}, \bibinfo{person}{Xun Zhou}, {et~al\mbox{.}}} \bibinfo{year}{2026}\natexlab{a}.
\newblock \showarticletitle{Ultramemv2: Memory networks scaling to 120b parameters with superior long-context learning}. In \bibinfo{booktitle}{\emph{ICLR}}.
\newblock


\bibitem[Huang et~al\mbox{.}(2025)]%
        {huangultra}
\bibfield{author}{\bibinfo{person}{Zihao Huang}, \bibinfo{person}{Qiyang Min}, \bibinfo{person}{Hongzhi Huang}, \bibinfo{person}{Yutao Zeng}, \bibinfo{person}{Defa Zhu}, \bibinfo{person}{Ran Guo}, {et~al\mbox{.}}} \bibinfo{year}{2025}\natexlab{}.
\newblock \showarticletitle{Ultra-Sparse Memory Network}. In \bibinfo{booktitle}{\emph{ICLR}}.
\newblock


\bibitem[Lai et~al\mbox{.}(2025)]%
        {lai2025exploring}
\bibfield{author}{\bibinfo{person}{Weijiang Lai}, \bibinfo{person}{Beihong Jin}, \bibinfo{person}{Jiongyan Zhang}, \bibinfo{person}{Yiyuan Zheng}, \bibinfo{person}{Jian Dong}, \bibinfo{person}{Jia Cheng}, \bibinfo{person}{Jun Lei}, {and} \bibinfo{person}{Xingxing Wang}.} \bibinfo{year}{2025}\natexlab{}.
\newblock \showarticletitle{Exploring Scaling Laws of CTR Model for Online Performance Improvement}. In \bibinfo{booktitle}{\emph{RecSys}}. \bibinfo{pages}{114--123}.
\newblock


\bibitem[Lample et~al\mbox{.}(2019)]%
        {lample2019large}
\bibfield{author}{\bibinfo{person}{Guillaume Lample}, \bibinfo{person}{Alexandre Sablayrolles}, \bibinfo{person}{Marc'Aurelio Ranzato}, \bibinfo{person}{Ludovic Denoyer}, {and} \bibinfo{person}{Herv{\'e} J{\'e}gou}.} \bibinfo{year}{2019}\natexlab{}.
\newblock \showarticletitle{Large memory layers with product keys}.
\newblock \bibinfo{journal}{\emph{NeurIPS}}  \bibinfo{volume}{32} (\bibinfo{year}{2019}).
\newblock


\bibitem[Li et~al\mbox{.}(2025)]%
        {li2025realizing}
\bibfield{author}{\bibinfo{person}{Dai Li}, \bibinfo{person}{Kevin Course}, \bibinfo{person}{Wei Li}, \bibinfo{person}{Hongwei Li}, \bibinfo{person}{Jie Hua}, \bibinfo{person}{Yiqi Chen}, \bibinfo{person}{Zhao Zhu}, \bibinfo{person}{Rui Jian}, \bibinfo{person}{Xuan Cao}, \bibinfo{person}{Bi Xue}, {et~al\mbox{.}}} \bibinfo{year}{2025}\natexlab{}.
\newblock \showarticletitle{Realizing Scaling Laws in Recommender Systems: A Foundation-Expert Paradigm for Hyperscale Model Deployment}.
\newblock \bibinfo{journal}{\emph{arXiv preprint arXiv:2508.02929}} (\bibinfo{year}{2025}).
\newblock


\bibitem[Li et~al\mbox{.}(2019)]%
        {fignn}
\bibfield{author}{\bibinfo{person}{Zekun Li}, \bibinfo{person}{Zeyu Cui}, \bibinfo{person}{Shu Wu}, \bibinfo{person}{Xiaoyu Zhang}, {and} \bibinfo{person}{Liang Wang}.} \bibinfo{year}{2019}\natexlab{}.
\newblock \showarticletitle{Fi-gnn: Modeling feature interactions via graph neural networks for ctr prediction}. In \bibinfo{booktitle}{\emph{CIKM}}. \bibinfo{pages}{539--548}.
\newblock


\bibitem[Lian et~al\mbox{.}(2018)]%
        {lian2018xdeepfm}
\bibfield{author}{\bibinfo{person}{Jianxun Lian}, \bibinfo{person}{Xiaohuan Zhou}, \bibinfo{person}{Fuzheng Zhang}, \bibinfo{person}{Zhongxia Chen}, \bibinfo{person}{Xing Xie}, {and} \bibinfo{person}{Guangzhong Sun}.} \bibinfo{year}{2018}\natexlab{}.
\newblock \showarticletitle{xdeepfm: Combining explicit and implicit feature interactions for recommender systems}. In \bibinfo{booktitle}{\emph{KDD}}. \bibinfo{pages}{1754--1763}.
\newblock


\bibitem[Lv et~al\mbox{.}(2025)]%
        {lv2025marm}
\bibfield{author}{\bibinfo{person}{Xiao Lv}, \bibinfo{person}{Jiangxia Cao}, \bibinfo{person}{Shijie Guan}, \bibinfo{person}{Xiaoyou Zhou}, \bibinfo{person}{Zhiguang Qi}, \bibinfo{person}{Yaqiang Zang}, \bibinfo{person}{Ben Wang}, {and} \bibinfo{person}{Guorui Zhou}.} \bibinfo{year}{2025}\natexlab{}.
\newblock \showarticletitle{MARM: Unlocking the Recommendation Cache Scaling-Law through Memory Augmentation and Scalable Complexity}. In \bibinfo{booktitle}{\emph{CIKM}}. \bibinfo{pages}{2022--2031}.
\newblock


\bibitem[Mao et~al\mbox{.}(2023)]%
        {mao2023finalmlp}
\bibfield{author}{\bibinfo{person}{Kelong Mao}, \bibinfo{person}{Jieming Zhu}, \bibinfo{person}{Liangcai Su}, \bibinfo{person}{Guohao Cai}, \bibinfo{person}{Yuru Li}, {and} \bibinfo{person}{Zhenhua Dong}.} \bibinfo{year}{2023}\natexlab{}.
\newblock \showarticletitle{FinalMLP: an enhanced two-stream MLP model for CTR prediction}. In \bibinfo{booktitle}{\emph{AAAI}}, Vol.~\bibinfo{volume}{37}. \bibinfo{pages}{4552--4560}.
\newblock


\bibitem[Mu et~al\mbox{.}(2026)]%
        {MDL}
\bibfield{author}{\bibinfo{person}{Shanlei Mu}, \bibinfo{person}{Yuchen Jiang}, \bibinfo{person}{Shikang Wu}, \bibinfo{person}{Shiyong Hong}, \bibinfo{person}{Tianmu Sha}, \bibinfo{person}{Junjie Zhang}, \bibinfo{person}{Jie Zhu}, \bibinfo{person}{Zhe Chen}, \bibinfo{person}{Zhe Wang}, {and} \bibinfo{person}{Jingjian Lin}.} \bibinfo{year}{2026}\natexlab{}.
\newblock \showarticletitle{MDL: A Unified Multi-Distribution Learner in Large-scale Industrial Recommendation through Tokenization}. In \bibinfo{booktitle}{\emph{KDD}}.
\newblock


\bibitem[Pan et~al\mbox{.}(2024)]%
        {pan2024ads}
\bibfield{author}{\bibinfo{person}{Junwei Pan}, \bibinfo{person}{Wei Xue}, \bibinfo{person}{Ximei Wang}, \bibinfo{person}{Haibin Yu}, \bibinfo{person}{Xun Liu}, \bibinfo{person}{Shijie Quan}, \bibinfo{person}{Xueming Qiu}, \bibinfo{person}{Dapeng Liu}, \bibinfo{person}{Lei Xiao}, {and} \bibinfo{person}{Jie Jiang}.} \bibinfo{year}{2024}\natexlab{}.
\newblock \showarticletitle{Ads recommendation in a collapsed and entangled world}. In \bibinfo{booktitle}{\emph{KDD}}. \bibinfo{pages}{5566--5577}.
\newblock


\bibitem[Qi et~al\mbox{.}(2025)]%
        {qi2025mtmixatt}
\bibfield{author}{\bibinfo{person}{Xianyang Qi}, \bibinfo{person}{Yuan Tian}, \bibinfo{person}{Zhaoyu Hu}, \bibinfo{person}{Zhirui Kuai}, \bibinfo{person}{Chang Liu}, \bibinfo{person}{Hongxiang Lin}, {and} \bibinfo{person}{Lei Wang}.} \bibinfo{year}{2025}\natexlab{}.
\newblock \showarticletitle{MTmixAtt: Integrating Mixture-of-Experts with Multi-Mix Attention for Large-Scale Recommendation}.
\newblock \bibinfo{journal}{\emph{arXiv preprint arXiv:2510.15286}} (\bibinfo{year}{2025}).
\newblock


\bibitem[Riquelme et~al\mbox{.}(2021)]%
        {riquelme2021scaling}
\bibfield{author}{\bibinfo{person}{Carlos Riquelme}, \bibinfo{person}{Joan Puigcerver}, \bibinfo{person}{Basil Mustafa}, \bibinfo{person}{Maxim Neumann}, \bibinfo{person}{Rodolphe Jenatton}, \bibinfo{person}{Andr{\'e} Susano~Pinto}, \bibinfo{person}{Daniel Keysers}, {and} \bibinfo{person}{Neil Houlsby}.} \bibinfo{year}{2021}\natexlab{}.
\newblock \showarticletitle{Scaling vision with sparse mixture of experts}.
\newblock \bibinfo{journal}{\emph{NeurIPS}}  \bibinfo{volume}{34} (\bibinfo{year}{2021}), \bibinfo{pages}{8583--8595}.
\newblock


\bibitem[Sadhukhan et~al\mbox{.}(2026)]%
        {sadhukhan2026stem}
\bibfield{author}{\bibinfo{person}{Ranajoy Sadhukhan}, \bibinfo{person}{Sheng Cao}, \bibinfo{person}{Harry Dong}, \bibinfo{person}{Changsheng Zhao}, \bibinfo{person}{Attiano Purpura-Pontoniere}, \bibinfo{person}{Yuandong Tian}, \bibinfo{person}{Zechun Liu}, {and} \bibinfo{person}{Beidi Chen}.} \bibinfo{year}{2026}\natexlab{}.
\newblock \showarticletitle{STEM: Scaling Transformers with Embedding Modules}. In \bibinfo{booktitle}{\emph{ICLR}}.
\newblock


\bibitem[Shan et~al\mbox{.}(2016)]%
        {deepcross}
\bibfield{author}{\bibinfo{person}{Ying Shan}, \bibinfo{person}{T~Ryan Hoens}, \bibinfo{person}{Jian Jiao}, \bibinfo{person}{Haijing Wang}, \bibinfo{person}{Dong Yu}, {and} \bibinfo{person}{JC Mao}.} \bibinfo{year}{2016}\natexlab{}.
\newblock \showarticletitle{Deep crossing: Web-scale modeling without manually crafted combinatorial features}. In \bibinfo{booktitle}{\emph{KDD}}. \bibinfo{pages}{255--262}.
\newblock


\bibitem[Shen et~al\mbox{.}(2024)]%
        {shen2024optimizing}
\bibfield{author}{\bibinfo{person}{Tingjia Shen}, \bibinfo{person}{Hao Wang}, \bibinfo{person}{Chuhan Wu}, \bibinfo{person}{Jin~Yao Chin}, \bibinfo{person}{Wei Guo}, \bibinfo{person}{Yong Liu}, \bibinfo{person}{Huifeng Guo}, \bibinfo{person}{Defu Lian}, \bibinfo{person}{Ruiming Tang}, {and} \bibinfo{person}{Enhong Chen}.} \bibinfo{year}{2024}\natexlab{}.
\newblock \showarticletitle{Optimizing sequential recommendation models with scaling laws and approximate entropy}.
\newblock \bibinfo{journal}{\emph{arXiv preprint arXiv:2412.00430}} (\bibinfo{year}{2024}).
\newblock


\bibitem[Song et~al\mbox{.}(2019)]%
        {song2019autoint}
\bibfield{author}{\bibinfo{person}{Weiping Song}, \bibinfo{person}{Chence Shi}, \bibinfo{person}{Zhiping Xiao}, \bibinfo{person}{Zhijian Duan}, \bibinfo{person}{Yewen Xu}, \bibinfo{person}{Ming Zhang}, {and} \bibinfo{person}{Jian Tang}.} \bibinfo{year}{2019}\natexlab{}.
\newblock \showarticletitle{Autoint: Automatic feature interaction learning via self-attentive neural networks}. In \bibinfo{booktitle}{\emph{CIKM}}. \bibinfo{pages}{1161--1170}.
\newblock


\bibitem[Swietojanski and Renals(2014)]%
        {lhuc}
\bibfield{author}{\bibinfo{person}{Pawel Swietojanski} {and} \bibinfo{person}{Steve Renals}.} \bibinfo{year}{2014}\natexlab{}.
\newblock \showarticletitle{Learning hidden unit contributions for unsupervised speaker adaptation of neural network acoustic models}. In \bibinfo{booktitle}{\emph{SLT Workshop}}. IEEE, \bibinfo{pages}{171--176}.
\newblock


\bibitem[Wang et~al\mbox{.}(2025)]%
        {wang2025scaling}
\bibfield{author}{\bibinfo{person}{Chunqi Wang}, \bibinfo{person}{Bingchao Wu}, \bibinfo{person}{Zheng Chen}, \bibinfo{person}{Lei Shen}, \bibinfo{person}{Bing Wang}, {and} \bibinfo{person}{Xiaoyi Zeng}.} \bibinfo{year}{2025}\natexlab{}.
\newblock \showarticletitle{Scaling transformers for discriminative recommendation via generative pretraining}. In \bibinfo{booktitle}{\emph{KDD}}. \bibinfo{pages}{2893--2903}.
\newblock


\bibitem[Wang et~al\mbox{.}(2017)]%
        {dcn}
\bibfield{author}{\bibinfo{person}{Ruoxi Wang}, \bibinfo{person}{Bin Fu}, \bibinfo{person}{Gang Fu}, {and} \bibinfo{person}{Mingliang Wang}.} \bibinfo{year}{2017}\natexlab{}.
\newblock \showarticletitle{Deep \& cross network for ad click predictions}. In \bibinfo{booktitle}{\emph{ADKDD}}. \bibinfo{pages}{1--7}.
\newblock


\bibitem[Wang et~al\mbox{.}(2021)]%
        {wang2021dcn}
\bibfield{author}{\bibinfo{person}{Ruoxi Wang}, \bibinfo{person}{Rakesh Shivanna}, \bibinfo{person}{Derek Cheng}, \bibinfo{person}{Sagar Jain}, \bibinfo{person}{Dong Lin}, \bibinfo{person}{Lichan Hong}, {and} \bibinfo{person}{Ed Chi}.} \bibinfo{year}{2021}\natexlab{}.
\newblock \showarticletitle{Dcn v2: Improved deep \& cross network and practical lessons for web-scale learning to rank systems}. In \bibinfo{booktitle}{\emph{WWW}}. \bibinfo{pages}{1785--1797}.
\newblock


\bibitem[Wang et~al\mbox{.}(2023)]%
        {wang2023augmenting}
\bibfield{author}{\bibinfo{person}{Weizhi Wang}, \bibinfo{person}{Li Dong}, \bibinfo{person}{Hao Cheng}, \bibinfo{person}{Xiaodong Liu}, \bibinfo{person}{Xifeng Yan}, \bibinfo{person}{Jianfeng Gao}, {and} \bibinfo{person}{Furu Wei}.} \bibinfo{year}{2023}\natexlab{}.
\newblock \showarticletitle{Augmenting language models with long-term memory}.
\newblock \bibinfo{journal}{\emph{NeurIPS}}  \bibinfo{volume}{36} (\bibinfo{year}{2023}), \bibinfo{pages}{74530--74543}.
\newblock


\bibitem[Zhai et~al\mbox{.}(2024)]%
        {zhai2024actions}
\bibfield{author}{\bibinfo{person}{Jiaqi Zhai}, \bibinfo{person}{Lucy Liao}, \bibinfo{person}{Xing Liu}, \bibinfo{person}{Yueming Wang}, \bibinfo{person}{Rui Li}, \bibinfo{person}{Xuan Cao}, \bibinfo{person}{Leon Gao}, \bibinfo{person}{Zhaojie Gong}, \bibinfo{person}{Fangda Gu}, \bibinfo{person}{Jiayuan He}, {et~al\mbox{.}}} \bibinfo{year}{2024}\natexlab{}.
\newblock \showarticletitle{Actions Speak Louder than Words: Trillion-Parameter Sequential Transducers for Generative Recommendations}. In \bibinfo{booktitle}{\emph{ICML}}. PMLR, \bibinfo{pages}{58484--58509}.
\newblock


\bibitem[Zhang et~al\mbox{.}(2024a)]%
        {zhang2024wukong}
\bibfield{author}{\bibinfo{person}{Buyun Zhang}, \bibinfo{person}{Liang Luo}, \bibinfo{person}{Yuxin Chen}, \bibinfo{person}{Jade Nie}, \bibinfo{person}{Xi Liu}, \bibinfo{person}{Shen Li}, \bibinfo{person}{Yanli Zhao}, \bibinfo{person}{Yuchen Hao}, \bibinfo{person}{Yantao Yao}, \bibinfo{person}{Ellie~Dingqiao Wen}, {et~al\mbox{.}}} \bibinfo{year}{2024}\natexlab{a}.
\newblock \showarticletitle{Wukong: Towards a Scaling Law for Large-Scale Recommendation}. In \bibinfo{booktitle}{\emph{ICML}}. PMLR, \bibinfo{pages}{59421--59434}.
\newblock


\bibitem[Zhang et~al\mbox{.}(2024b)]%
        {dpn}
\bibfield{author}{\bibinfo{person}{Hengyu Zhang}, \bibinfo{person}{Junwei Pan}, \bibinfo{person}{Dapeng Liu}, \bibinfo{person}{Jie Jiang}, {and} \bibinfo{person}{Xiu Li}.} \bibinfo{year}{2024}\natexlab{b}.
\newblock \showarticletitle{Deep pattern network for click-through rate prediction}. In \bibinfo{booktitle}{\emph{SIGIR}}. \bibinfo{pages}{1189--1199}.
\newblock


\bibitem[Zhang et~al\mbox{.}(2023)]%
        {zhang2023parallel}
\bibfield{author}{\bibinfo{person}{Jingrong Zhang}, \bibinfo{person}{Akira Naruse}, \bibinfo{person}{Xipeng Li}, {and} \bibinfo{person}{Yong Wang}.} \bibinfo{year}{2023}\natexlab{}.
\newblock \showarticletitle{Parallel top-k algorithms on gpu: A comprehensive study and new methods}. In \bibinfo{booktitle}{\emph{SC}}. \bibinfo{pages}{1--13}.
\newblock


\bibitem[Zhang et~al\mbox{.}(2019)]%
        {zhang2019deep}
\bibfield{author}{\bibinfo{person}{Shuai Zhang}, \bibinfo{person}{Lina Yao}, \bibinfo{person}{Aixin Sun}, {and} \bibinfo{person}{Yi Tay}.} \bibinfo{year}{2019}\natexlab{}.
\newblock \showarticletitle{Deep learning based recommender system: A survey and new perspectives}.
\newblock \bibinfo{journal}{\emph{CSUR}} \bibinfo{volume}{52}, \bibinfo{number}{1} (\bibinfo{year}{2019}), \bibinfo{pages}{1--38}.
\newblock


\bibitem[Zhao et~al\mbox{.}(2023)]%
        {zhao2023survey}
\bibfield{author}{\bibinfo{person}{Wayne~Xin Zhao}, \bibinfo{person}{Kun Zhou}, \bibinfo{person}{Junyi Li}, \bibinfo{person}{Tianyi Tang}, \bibinfo{person}{Xiaolei Wang}, \bibinfo{person}{Yupeng Hou}, \bibinfo{person}{Yingqian Min}, \bibinfo{person}{Beichen Zhang}, \bibinfo{person}{Junjie Zhang}, \bibinfo{person}{Zican Dong}, {et~al\mbox{.}}} \bibinfo{year}{2023}\natexlab{}.
\newblock \showarticletitle{A survey of large language models}.
\newblock \bibinfo{journal}{\emph{arXiv preprint arXiv:2303.18223}} \bibinfo{volume}{1}, \bibinfo{number}{2} (\bibinfo{year}{2023}).
\newblock


\bibitem[Zhong et~al\mbox{.}(2024)]%
        {zhong2024memorybank}
\bibfield{author}{\bibinfo{person}{Wanjun Zhong}, \bibinfo{person}{Lianghong Guo}, \bibinfo{person}{Qiqi Gao}, \bibinfo{person}{He Ye}, {and} \bibinfo{person}{Yanlin Wang}.} \bibinfo{year}{2024}\natexlab{}.
\newblock \showarticletitle{Memorybank: Enhancing large language models with long-term memory}. In \bibinfo{booktitle}{\emph{AAAI}}, Vol.~\bibinfo{volume}{38}. \bibinfo{pages}{19724--19731}.
\newblock


\bibitem[Zhou et~al\mbox{.}(2019)]%
        {zhou2019deep}
\bibfield{author}{\bibinfo{person}{Guorui Zhou}, \bibinfo{person}{Na Mou}, \bibinfo{person}{Ying Fan}, \bibinfo{person}{Qi Pi}, \bibinfo{person}{Weijie Bian}, \bibinfo{person}{Chang Zhou}, \bibinfo{person}{Xiaoqiang Zhu}, {and} \bibinfo{person}{Kun Gai}.} \bibinfo{year}{2019}\natexlab{}.
\newblock \showarticletitle{Deep interest evolution network for click-through rate prediction}. In \bibinfo{booktitle}{\emph{AAAI}}, Vol.~\bibinfo{volume}{33}. \bibinfo{pages}{5941--5948}.
\newblock


\bibitem[Zhou et~al\mbox{.}(2018)]%
        {zhou2018deep}
\bibfield{author}{\bibinfo{person}{Guorui Zhou}, \bibinfo{person}{Xiaoqiang Zhu}, \bibinfo{person}{Chenru Song}, \bibinfo{person}{Ying Fan}, \bibinfo{person}{Han Zhu}, \bibinfo{person}{Xiao Ma}, \bibinfo{person}{Yanghui Yan}, \bibinfo{person}{Junqi Jin}, \bibinfo{person}{Han Li}, {and} \bibinfo{person}{Kun Gai}.} \bibinfo{year}{2018}\natexlab{}.
\newblock \showarticletitle{Deep interest network for click-through rate prediction}. In \bibinfo{booktitle}{\emph{KDD}}. \bibinfo{pages}{1059--1068}.
\newblock


\bibitem[Zhou et~al\mbox{.}(2024)]%
        {tin}
\bibfield{author}{\bibinfo{person}{Haolin Zhou}, \bibinfo{person}{Junwei Pan}, \bibinfo{person}{Xinyi Zhou}, \bibinfo{person}{Xihua Chen}, \bibinfo{person}{Jie Jiang}, \bibinfo{person}{Xiaofeng Gao}, {and} \bibinfo{person}{Guihai Chen}.} \bibinfo{year}{2024}\natexlab{}.
\newblock \showarticletitle{Temporal Interest Network for User Response Prediction}. In \bibinfo{booktitle}{\emph{WWW}}. \bibinfo{pages}{413--422}.
\newblock


\bibitem[Zhu et~al\mbox{.}(2025a)]%
        {zhu2025rankmixer}
\bibfield{author}{\bibinfo{person}{Jie Zhu}, \bibinfo{person}{Zhifang Fan}, \bibinfo{person}{Xiaoxie Zhu}, \bibinfo{person}{Yuchen Jiang}, \bibinfo{person}{Hangyu Wang}, \bibinfo{person}{Xintian Han}, \bibinfo{person}{Haoran Ding}, \bibinfo{person}{Xinmin Wang}, \bibinfo{person}{Wenlin Zhao}, \bibinfo{person}{Zhen Gong}, {et~al\mbox{.}}} \bibinfo{year}{2025}\natexlab{a}.
\newblock \showarticletitle{Rankmixer: Scaling up ranking models in industrial recommenders}. In \bibinfo{booktitle}{\emph{CIKM}}. \bibinfo{pages}{6309--6316}.
\newblock


\bibitem[Zhu et~al\mbox{.}(2025b)]%
        {zhu2025addressing}
\bibfield{author}{\bibinfo{person}{Yongxin Zhu}, \bibinfo{person}{Bocheng Li}, \bibinfo{person}{Yifei Xin}, \bibinfo{person}{Zhihua Xia}, {and} \bibinfo{person}{Linli Xu}.} \bibinfo{year}{2025}\natexlab{b}.
\newblock \showarticletitle{Addressing representation collapse in vector quantized models with one linear layer}. In \bibinfo{booktitle}{\emph{ICCV}}. \bibinfo{pages}{22968--22977}.
\newblock


\end{thebibliography}

\appendix

\end{document}